\documentclass[12pt]{article}
\usepackage{ifpdf}
\usepackage{graphicx,natbib,amssymb,amsmath}
\usepackage{latexsym,theorem}
\usepackage{euscript}
\usepackage[english]{babel}

\theoremstyle{plain}
\newtheorem{theorem}{Theorem}

\newtheorem{proposition}[theorem]{Proposition}

{\theorembodyfont{\upshape}

\newtheorem{algorithm}[theorem]{Algorithm}

\newtheorem{remark}[theorem]{Remark}

\newenvironment{Proof}{\small{\bf Proof.}}{\hfill$\Box$\normalsize
\bigskip}

\newcommand{\EE}{{\mathbb E}} %
\newcommand{\PP}{{\mathbb P}} %

\newcommand{\NN}{{\mathbb N}}

\def\text#1{\quad\mbox{#1}\quad} 
\def\defegal{\triangleq}

\def\F{\mathcal{F}}

\newenvironment{keywords}{\small{\bf
    Key Words:}}{\hfill\normalsize\\ \bigskip}

\title{
Gas Fire Power Plant Management Through Numerical Approximation of Spark Spread Options}

\author{Babacar Seck$^{1}$\\
Assistant Professor of Mathematical Finance and Actuarial Science\\
Department of Mathematics, University of Bahrain\\
Sakhir, P.O. Box 32038\\
Kingdom of Bahrain \\[3mm]
Anas Abdullah$^{2}$\\
Associate Professor of Mathematical Finance and Actuarial Science\\
Department of Mathematics \& Statistics, McMaster University\\
Hamilton Hall, Room 408, 1280 Main Street West\\
Hamilton, ON L8S 4K1, Canada \\[3mm]
{}$^{1}$bseck@uob.edu.bh,\;{}$^{2}$anasabdallah@mcmaster.ca\\[3mm]
}

\begin{document}
\maketitle

\vspace{0.5cm}
\begin{abstract}
Cross-commodity valuation approaches to value gas fire power plants are well studied in the literature. Hence, the value of the gas fire power plant is identical to the value of a spark spread option wherein the underlying are electricity and gas with a strike price assimilated to operating and maintenance costs. Power and fuels spot prices account for uncertain futures cash-flows for power-plant generator owners. For instance, for gas-fired turbine plant, spot prices of electricity and gas determine the random cash-flows of the power-plant. Other than the spot prices, the valuation of such plant involves among other deterministic cost the plant heat rate and operating costs. Recently, the cost of emissions is considered into the valuation to tackle environmental issues. Given some simplifications in the plant cash-flow modelling, the value of such plant can either be expressed as the price of  i) a cross-commodity option or ii) the price of a real option. Here, we focus on cross-commodity option valuation approach where the value of the power plant is approached as the value of a spark spread option. When spot prices of the underlying commodities are log-normal, closed formulae or approximations can be obtained using Kirk’s approximation. Naturally, the spot price of electricity and gas present spikes due to seasonality among other factors. However, in that case it is not possible to get a closed formula for the spark spread option. In this paper we explore possibilities to approximate spark spread options when spot prices fall into a class of jump diffusion processes.
\end{abstract}


\section{Introduction}
A gas-fired power plant is a thermal station that burns natural gas to electricity. Globally, almost a quarter of the entire electricity consumed in the word is generated from gas-fire power plant. This contribution of gas-fired power plant in the electricity production is highly significant for some countries. For instance in the Kingdom of  Bahrain almost all the electricity consumed is from gas-fired power plant. Proper management of gas-fired power plant is therefore critical and may required to take into account different aspects linked to operating the plant itself, the integration of the plant into a grid with a growing share of  renewable energy, and the necessity of environmental regulations. The management of operating the plant should integrate the fuel supply, the plant performance monitoring and optimization, the maintenance and reliability, environmental control and compliance, safety and risk, the electricity production, as so one. Here, we will focus mainly on the production aspect without considering the integration of the production unit into a grid system.  Therefore, the variables we will be considering are  the operating cost, the heat rate, and the prices of the gas and electricty in the context of a specific power pool and energy hub market. Then, managing the plant  cashflows in and out is reduced to valuing the spreads of a cross-commodity option.

Power and fuels spot prices account for uncertain futures cash-flows for power-plant generator owners.
For instance, for gas-fired turbine plant, spot prices of electricity and gas determine the random cash-flows of the power-plant.
Other than the spot prices, the valuation of such plant involves among other deterministic cost the plant heat rate and operating costs. The cost of emissions can be integrated  into the valuation, as in~\cite{Rene2012} to tackle environmental issues.  Given some simplifications in the 
plant cash-flow modelling, the value of such plant can either be expressed as the price of a i)cross-commodity option  or ii) the price of a real option. 
Here, we focus on cross-commodity option valuation approach
where the value of the power plant is approached as the value of a spark spread option. Seminal works on valuation of spread option are~\cite{Margrabe1978},~\cite{Carmona-Durrleman2003}.
The problem we address is to price  spark spread options based on forward curves and spot price models.
Let  $F_e(\tau,\tau_1,\tau_2)$ in (\$/MWh) be the  electricity futures price for a delivery period $\big[\tau_1,\tau_2\big]$, with $\tau<\tau_1$. Recall that, the settlement of  such futures contract should be defined with respect to the average spot price as spot prices of electricity are quoted in a hourly basis. Assume that $S_e(\tau)$ is the daily spot 
price of electricity.  Let  $F_g(\tau,\tau_1,\tau_2)$ in (\$/GJ) be the  gas futures price over the same delivery period. 
Note that, the settlement of gas futures contract are usually based on monthly delivery periods. Define $S_g(\tau)$ the daily spot price of gas. The heat rate or plant efficiency, denoted $r_g$ is the amount of GJ of natural gas necessary to acquire one MWh of electricity. We assume that $r_g$ is not  time independent and is a characteristic of the power plant. This latter is different from the market heat rate. The market heat rate is the ratio of the electricity price to the
 natural gas price. The operating and maintenance cost necessary to run the plant is also constant and is denoted $K$. Assume that, the gas power plant runs in two regimes (shut down or run)  and the cost of changing regime is null. A simple arbitrage reasoning leads to the valuation formulae for such gas fired power plant. For instance, it is profitable to run such plant when 
$\displaystyle F_e(\tau,\tau_1,\tau_2)> r_gF_g(\tau,\tau_1,\tau_2) + K$. Hence  the value of the power plant  at time $t<\tau$, 
$V(t)$ can be expressed as 

$$V(t)=e^{-r_f(t-\tau)}\Big(\displaystyle F_e(\tau,\tau_1,\tau_2)-r_gF_g(\tau,\tau_1,\tau_2) - K\Big),$$ 
where $r_f$ represents the risk-free interest rate. 
If $\displaystyle F_e(\tau,\tau_1,\tau_2)\leq  r_gF_g(\tau,\tau_1,\tau_2) + K$, it is not worth to operate the plant and $V(t)=0$.
Then,  the value of such gas power plant over the (daily) delivery period $[\tau_1;\tau_2]$ is

\begin{equation}
\label{price_1}
V(t)=
e^{-r_f(t-\tau)}\EE\Big[\Big(\displaystyle F_e(\tau,\tau_1,\tau_2)-r_gF_g(\tau,\tau_1,\tau_2) - K\Big)^{+}\mid \mathcal{F}_t \Big],
\quad \forall t<\tau,
\end{equation} 
where $x^{+}$ stands for the positive part of $x$, i.e. $x^+=x$ if $x>0$ and $x^+=0$ if $x\leq 0$ and $\mathcal{F}_t$ is the information available at day $t$. Hence, the expected value is taken over the joint distribution of the two forward contracts which result from the two spot prices. Also, the granularities of the two spot prices are not the same since electricity is quoted hourly while gas are quoted one day ahead. This gives an indication on the difficuties of finding closed formulae or numerical solutions for $V(t)$. Note that a risk-neutral pricing approach is not realistic since the energy markets are highly illiquid.   The pricing formulae in~\eqref{price_1}, is known in the litterature as the Actuarial formulae of the gas fired power plant.

Upper and lower bound of the spark spread call option $V(t)$ can be derived using formulas in~\cite{Deng2001}, when $K=0$. In the wholesale market, electricity and gas prices are quoted for some future delivery period.
For the sake of simplicity, we assume that the spot prices $S_e$ and $S_g$ are  observed in the market at time $t$, for electricity and gas, respectively. Recall that the electricity spot price are quoted hourly while the gas spot price is  quoted one-day ahead. This paper is organized as following. In Section~\ref{sec:ModelDiff}, we adapt additive diffusion models used in weather forcast, to model the spot prices of gas and electricity. Preliminary numerical simulations has shown promising results in using such model, compared to previously used models.  In Section~\ref{sec:PricingMethods}, we discuss the prcing methodology of speark spread option, which corresponds to the value cashflows of a gas fired power plant.
Concluding remarks are given in Section~\ref{sec:Concl}.

\section{Model of spot prices diffusions for electricity and gas}
\label{sec:ModelDiff}
Let $(\Omega,\F,\PP)$ be a probability space equipped with a filtration $\F_t$.
We assume that, spot prices are derived from arithmetic mean-reverting long term models with jumps.
Following~\cite{LuciaSchwartz2002},~\cite{Benth2007} and~\cite{Benth2008}, we consider the electricity spot price process defined by:
\begin{equation}
\label{eq_spot1}
S_e(t)=\Lambda_e(t)+X_e(t) + Y_e(t),
\end{equation}
where  $\Lambda_e: [0,\mathcal{T}\,]\mapsto (0;+\infty)$ stands for a deterministic seasonal spot prices level.  $X_e(t)$ and $Y_e(t)$ are the solutions of the following stochastic differential equations:
\begin{equation}
\label{eq_spot2}
\mathrm{d}X_e(t)= - \alpha_e(t)X_e(t)\mathrm{d}t+\sigma_e(t)\mathrm{d}B_e(t)
\end{equation}
 and 
\begin{equation}
 \mathrm{d}Y_e(t)= - \beta_e(t)Y_e(t)\mathrm{d}t+\eta_e(t)\mathrm{d}I_{e}(t).
\end{equation}
 Here, $B_e$ is  a standard Brownian motion  and $I_e$ is a pure jump semimartingale
 independent increments process. Next, we shall identify $I_e$ as a L\'evy process. The process $Y_e$ is a zero-mean reverting processes responsible for
the spikes or large deviations which revert at a fast rate $\beta_e(t)>0$ while $X_e$ is a zero-mean
reverting processes that account for the normal variations in the spot price evolution with
mean-reversion $\alpha_e(t)>0$. The functions $\alpha_e,\;\beta_e,\;\sigma_e$ and $\eta_e$ are continuous functions. 
The equation~\eqref{eq_spot1}  defines a two factors mean-reverting stochastic process with jump.
 Arithmetic processes as defined in~\eqref{eq_spot1} gained popularity in modelling daily temperature. 
 See~\cite{Benth2012} for a complete review on additive models with jump for spot prices of energy commodity.
 Gas spot prices\footnote{Gas spot prices, usually refers to the short term delivery price.} present similarity with electricity spot prices with sudden spikes during period of high demand or low storage.  In the model~\eqref{eq_spot1}, negative spot prices are possible, which had happened in some electricity markets. However, such behavior has not been observed in the gas market. For that reason, we consider the spot prices of gas as a geometric model with jump in the form:
 \begin{equation}
\ln \Big(S_g(t)\Big)=\ln\Big(\Lambda_g(t)\Big)+X_g(t) + Y_g(t),
\end{equation}
where  $\Lambda_g: [0,\mathcal{T}\,]\mapsto (0;+\infty)$ stands for a deterministic seasonal gas spot price level.  $X_g(t)$ and $Y_g(t)$ are the solutions of the following stochastic differential equations:
\begin{equation}
\label{eq_spot2_gas}
\mathrm{d}X_g(t)= - \alpha_g(t)X_g(t)\mathrm{d}t+\sigma_g(t)\mathrm{d}B_g(t)
\end{equation}
 and 
\begin{equation}
 \mathrm{d}Y_g(t)= - \beta_g(t)Y_g(t)\mathrm{d}t+\eta_g(t)\mathrm{d}I_{g}(t),
\end{equation}
where $B_g$ is a standard Brownian motion and $I_g$ a L\'evy process. 
The  above spot price model is a generalization of  the so-called  Schwartz model with jump. In the perspective of valuing the spark spread option defined in~\eqref{price_1}, dependance between electricity and gas spot prices need to be addressed.
Assuming that  the normal variations in the spot prices models are identical, i.e. $X_g=X_e$, then  the auto-correlation functions and the cross-correlation functions between the two spot prices can be derived using formulas obtained in~\cite{Frikha2013}. However, it is not realistic to make such assumption as spot prices of gas and electricity are widely different.
The models correlation will be discussed later. Recently, a basic correlation analysis of the spot price of electricity and gas have revealed 
a limited linear correlation, in the Alberta's electricity market. The coefficient of correlation depends on the seasonality and the average price spot price of the electricity. We keep in mind that the delivery time frame scale of gas is at least monthly, while electricity can be quoted for a delivery for the next day. 
 
 \section{Pricing methodologies of spark spread options}
\label{sec:PricingMethods}
Different approaches of pricing spark spread options exist in the literature. As usual, the pricing methods presented 
depend on the underlying spot price model (model with jumps or not, short term or long term models) of the electricity and gas. Next, we present a brief review of those methods of pricing. A comparison based on numerical simulations will also be presented.  \cite{Cassano2011} applied the Margrabe approach to model market heat rate and then valued a gas-fired turbine plant. 
The approaches  developed are Least Square Monte Carlo simulation and linear regression method taking into account  4 levels of  plant operating modes. The operating characteristics, demand, merit-order
are also considered in their modeling. Carmona, Toulon and Schwarz  valued  gas-fired plant or clean spread option which take into account carbon emissions, with  four source of randomness:  the  spot prices of fuels  (electricity and gas), the emission market  and demand (\emph{c.f.}~\cite{Rene2012}). Then, they solved a set of 4 stochastic differential equations. In this settings,  the characteristics of the gas-fired plant are not taken into account in the valuation principle.
In \cite{Hepperger2010},  european style option pricing methods for time dependant Hilbert space spot prices with jump-diffusions using two
numerical approaches are developed: a partial Integro-Differential equation methods and a dimension reduction methods based on 
Karhunen-Lo\`{e}ve expansion. However, for basket option which is considered in  this paper, the PIDE method yields to comparable method to
Monte Carlo simulation depending on the correlations on the underlying asset. In fact, a basket option of two or three assets can be expressed
as a spark spread option (or a clean spark option) where the strike represents the operating cost of converting the fuel to electricity and
the eventual third underlying could express the prices of emissions.

Recall that   the value of a gas power plant over the (monthly) delivery period $[\tau_1;\tau_2]$ can be expressed as:
\begin{equation}
\label{Eq:General}
V(t)=
e^{-r_f(t-\tau)}\EE\Big[\Big(\displaystyle F_e(\tau,\tau_1,\tau_2)-r_gF_g(\tau,\tau_1,\tau_2) - K\Big)^{+}\mid \mathcal{F}_t \Big],
\quad \forall t<\tau.
\end{equation}
 Independently of choices for the dynamics of the prices of electricity and gas, extended formulae of the put-call parity may allow to find upper and lower bound of $V(t)$, when $K=0$. This result is obtained in~\cite{Deng2001} and states that
\begin{equation}
\label{eq:Bounds}
 e^{-r_f(t-\tau)} \Big(\displaystyle F_e(\tau,\tau_1,\tau_2)-r_gF_g(\tau,\tau_1,\tau_2) \Big)^{+} \leq V(t) \leq 
 e^{-r_f(t-\tau)} F_e(\tau,\tau_1,\tau_2).
\end{equation}
The equation~\eqref{eq:Bounds} will be helpfull when closed formulae of $V(t)$ can not be obtained.
Assume that, the valuation of the spark-spread option is based on the day-ahead electricity and gas prices. Hence, 
$F_e(t,\tau_1,\tau_2)$ and $F_g(t,\tau_1,\tau_2)$ can be rewritten in the form
$$
F_e(t,\tau_1,\tau_2)=\displaystyle \frac{1}{\tau_2 - \tau_1}\sum_{t=\tau_1}^{\tau_2} e^{-r_t} S_e(t) \quad\mbox{and}\quad
F_g(t,\tau_1,\tau_2)=\displaystyle \frac{1}{\tau_2 - \tau_1}  \sum_{t=\tau_1}^{\tau_2} e^{-r_t}  S_g(t). 
$$
Then, possessing a spark-spread call option is equivalent to
owning a power plant with operational flexibility (if operational costs and operational
constraints are disregarded). In this case, the spark-spread option replicates a power plant under
the assumption that the plant can be operated daily in peak hours at no cost besides the natural gas
consumption. Under these assumptions, closed formulas are obtained in~\cite{Maribu2007} when $S_e$ and $S_g$
are modelled by a mean-reverting stochastic volatility process.

\subsection{Price approximation for spot Merton's jump models}
Following the so-called Merton's jump diffusion process, assume
that the underlying $S_e$ and $S_g$ are given by the stochastic differential equation of the form:
\begin{equation}
\label{MertonJump}
\frac{\mathrm{d}S_i(t)}{S_i(t-)}=(r - q_i - \lambda_i \kappa_i)\mathrm{d}t +\sigma_i \mathrm{d}B_i(t) + \big( e^{J_i(t)} - 1  \big)\mathrm{d}N_i(t),
\quad i=e,g,
\end{equation} 
where $N_e$ and $N_g$ are two independent Poisson processes with intensity $\lambda_e$ and $\lambda_g$, respectively.
Also, the Brownian motions $B_e$ and $B_g$ are assumed to be independent. Finally, the $(J_e(t))_{(t\geq 0)}$ and  $(J_g(t))_{(t\geq 0)}$
are independent sequences of independent Gaussian random variables with means $m_i$ and variances $s_i^2$. By applying the extended
It\^{o}'s Lemma for jump diffusion processes, the integrated prices dynamic is:
\begin{equation}
S_i(T)=S_i(0)\exp\!\!\Big(  (r-q_i-\sigma_i^2/2 -\lambda_i\kappa_i) T + \sigma_i B_i + \sum_{k=1}^{N(T)}J_i(k)\Big).
\end{equation}
By conditioning with respect to the Poisson process first, the problem of finding the price $V$ is reduced to pricing a simple spark-spread option with log-normal underlying distributions. Then  given the $N_i(T)$, the random variables $S_i(T)$ are log-normal. Based on this observations,  an approximation $\widehat{V}$ of the value of the spark-spread option $V$ can be derived using
 the formula in~\cite{Carmona-Durrleman2003}.
 
 \begin{proposition}[\cite{Carmona-Durrleman2003}]Assume that the stochastic processes $S_e$ and $S_g$ are driven by the above Merton's jump model.
 Assume that the value of gas fire plant $V_d$ is priced day-ahead.
 Assume that $\hat{p}$ is the price approximation of the spread option $p$ given by
 \begin{equation}
 p\defegal\EE\Big[  e^{-rT}(S_e(T) - S_g(T) -K)^{+}\Big]
 \end{equation}
where the expectation computed under a risk-neutral probability.
  If we set $\mu_i=e^{(m_i +s_i^2/2)}-1$,   an approximation of  the daily value of the spread option $\widehat{V}_d$ is:
 \begin{equation}
 \label{priceinfinite}
 \widehat{V}_d =\sum_{i=0}^{\infty}\sum_{j=0}^{\infty} e^{-(\lambda_1 +\lambda_2)T}
 \frac{(\lambda_1 T)^i(\lambda_2 T)^j}{i!j!} \hat{p}(\tilde{x_e},\tilde{x_g},\tilde{\sigma_e},\tilde{\sigma_g},\tilde{\rho}),
 \end{equation} 
 where
 \begin{equation}
  \tilde{x_e}=s_e(0)e^{\lambda_e\mu_e T +i(m_e +s_e^2/2)}, \quad  \tilde{x_g}=s_g(0)e^{\lambda_g\mu_g T +j(m_g +s_g^2/2)}
 \end{equation}
 and 
  \begin{equation}
 \tilde{\sigma_e}=\sqrt{\sigma_e + is_e^2/T}, \quad  \tilde{\sigma_g}=\sqrt{\sigma_g + js_g^2/T}\quad\mbox{and}\quad 
\tilde{\rho} = \frac{\rho\sigma_e\sigma_g}{ \tilde{\sigma_e} \tilde{\sigma_g}}.
 \end{equation}
\label{Prop1}
 \end{proposition} 

 There are a certain number of considerations regarding  Proposition~\ref{Prop1}. First, the correlation between the spot electricity and gas prices are easily observable in the market. Second, a risk neutral approach can not be applied since the energy market are incomplete markets, for several reasons. From a numerical point of view, the price approximation of the spark spread option, depends on another approximation based on a risk-neutral expected value which depends on the paths of $S_i$, $i=e,g$, which makes the simulations quite chanllenging. As far as we are aware of, there is no simulations available in the litterature that are confirming the soundness of the approximation in the Proposition~\ref{Prop1}. 

\subsubsection{Price approximation using  double integrals}
For simplicity, there is no difference between the risk neutral probability and the historical probability. Hence, 
 $p$ can be approximated by  using the double integral:
 \begin{equation}
 \label{double_int}
 p=\int\Big( \int (x_e - x_g -K)^{+ }f_{e,T\mid S_g(T)=x_g}(x_e)\mathrm{d}x_e  \Big)f_{g,T}(x_g)\mathrm{d}x_g,
 \end{equation}
 where $f_{g,T}$ is the density function of the gas spot price at maturity $S_g(T)$ and $f_{e,T\mid S_g(T)=x_g}$
 is the conditional density function of the electricity price at maturity given that the gas spot price is equal to $x_g$ at that time.
 The approximation price $\tilde{p}$ can be computed by approximating the above double integral. Another way to approximate $p$ is to determinate 
$ \widehat{V}_d$. Both approximations  can be done through simulations only. In the following, we propose a method of approximating $ \widehat{V}_d$.  Consider the following 
 algorithm, which approximates the infinite series~\eqref{priceinfinite}.

 \begin{algorithm}Approximation of  $\widehat{V}_d$.
 \begin{enumerate}
 \item Initialization: 
 \begin{enumerate}
\item  Compute the approximation  $\tilde{p}^{0,0}$:
 $$\tilde{p}^{0,0}=\tilde{p}^{0,0}(\tilde{x_e}^{0},\tilde{x_g}^{0},\tilde{\sigma_e}^{0},\tilde{\sigma_g}^{0},\tilde{\rho}^{0,0}) \quad\mbox{and set}\quad
 \widehat{v}^{0,0}_{d}=
 e^{-(\lambda_1 +\lambda_2)T}\tilde{p}^{0,0}.
 $$
\item Set  the approximation error relative to the day ahead spread option $\mathcal{E}_d(0,0)=0.0005$.
\end{enumerate}
 \item Repeat
 \begin{enumerate}
 \item Approximate the double integral~\eqref{double_int}  up to iteration $k,l$ to obtain 
 $$\tilde{p}^{k,l}=\tilde{p}^{k,l}(\tilde{x_e}^{k},\tilde{x_g}^{l},\tilde{\sigma_e}^{k},\tilde{\sigma_g}^{l},\tilde{\rho}^{k,l}),\; 
 \widehat{v}^{k,l}_{d}=e^{-(\lambda_1 +\lambda_2)T}  \frac{(\lambda_1 T)^k(\lambda_2 T)^l}{k!l!}\tilde{p}^{k,l}  \mbox{ and } $$ 
  $\widehat{V}^{k,l}_{d}=\displaystyle \sum_{i=0}^{k}\sum_{j=0}^{l} \widehat{v}^{k,l}_{d}
 \tilde{p}^{k,l}. $

 \item Compute $\mathcal{E}_d(k,l)=\Big | \widehat{V}^{k,l}_{d} - \widehat{V}^{k-1,l-1}_{d}  \Big |$.
 \end{enumerate}
 \item Until $ \mathcal{E}_d(k,l)\leq \beta$, where $\beta$ is a small positive number and $k,l\in\NN$ .
 \end{enumerate}
\label{Algo:First}
 \end{algorithm}
\begin{remark}
The algorithm tells us that we will stop when the price approximate is stable. The convergence result depends on the properties of the $\widehat{v}^{k,l}_{d}$, since at each step of the simulation $\tilde{p}^{k,l}$ is real number. We will study in details the properties of  $\widehat{v}^{k,l}_{d}$.
\end{remark}
\noindent \textbf{Convergence of the Algorithm~\ref{Algo:First}}
 It is traighforward to see that
$$
\displaystyle \frac{\widehat{v}^{k+1,l+1}_{d}}{\widehat{v}^{k,l}_{d}}=
\frac{\lambda_1 \lambda_2 T^2}{(k+1)(l+1)} \frac{\tilde{p}^{k+1,l+1}}{\tilde{p}^{k,l}}
$$
Denote $D=\displaystyle\frac{\tilde{p}^{k+1,l+1}}{\tilde{p}^{k,l}}$. Then the operator $\widehat{v}^{k,l}_{d}$ is coercive, which guarantees the convergence of the algorithm.

\subsubsection{Price approximation from closed formula without jumps}
Besides the numerical difficulties in putting in practice the above algorithm, one of the downside of this algorithm is to find a benchmark.  A method to benchmark our algorithm is to find the corresponding closed formula of the spark spread option for log-OU without jumps. Then, we will attempt to see how far we can be from the closed formula when we control the frequency and the severity of the jump. Without jumps, the value of $p(\tilde{x_e},\tilde{x_g},\tilde{\sigma_e},\tilde{\sigma_g},\tilde{\rho},K)$ is given by Kirk's formula in~\cite{Kirk1995}:
\small{
\begin{equation}
\label{Kirk1}
p(x_e,x_g,\sigma_e,\sigma_g,\rho,K)=x_g\Phi\left( \frac{ \displaystyle  \ln\left( \frac{x_g}{x_e+Ke^{-r_f\delta_t}}\right)}{\sigma} + \frac{\sigma}{2}\right)-
\left(x_e +Ke^{-r_f}\delta_t \right)\Phi\left( \frac{ \displaystyle  \ln\left( \frac{x_g}{x_e+Ke^{-r_f\delta_t}}\right)}{\sigma} - \frac{\sigma}{2}     \right),
\end{equation} }
 where $\delta_t=T-t$ and 
\begin{equation}
\label{Kirk_sigma}
\sigma=\displaystyle \sqrt{\sigma_g^2 - 2\rho\sigma_e\sigma_g  \frac{x_e}{x_e+Ke^{-r_f\delta_t}} +\sigma_e^2\left(\frac{x_e}{x_e+Ke^{-r_f\delta_t}}\right)^2 }.
\end{equation}
Then substitute $\hat{p}(\tilde{x_e},\tilde{x_g},\tilde{\sigma_e},\tilde{\sigma_g},\tilde{\rho}, K)$ in the equation~\eqref{priceinfinite} to obtain an approximation of the spark spread option.
The approximations we have been discussing so far, are for options of the spot prices. It might be more relaistic, to word directly on underlying which are forwrd contract due the quotation and the delivery logistics in the gas markets. 

\subsection{Price approximation from assumptions on electricity and gaz spot prices correlations}
Based on the historical distribution of the electricity and gas prices, we can look at possible dependence. Depending on the seasonality, it is possible to have a linear approximation of the electricity spot price based on the gas price or vice versa. Then, we can write $F_{e}(t)=a(t)F(S_g(t)) + b(t)$, where $F(\cdot)$ is a real value function. By doing so, the pricing of the spark spread option is reduced to the pricing of a single underling asset, says $S_g$. Then the pricing equation \eqref{Eq:General} becomes
\begin{equation}
\label{Eq:General3}
V(t)=
e^{-r_f(t-\tau)}\EE\Big[\Big(\displaystyle a(t)F(S_g(t)) + b(t) - r_gF_g(\tau,\tau_1,\tau_2) - K\Big)^{+}\mid \mathcal{F}_t \Big],
\quad \forall t<\tau.
\end{equation}
Now, write $\widehat{F}_{t,\tau}(S_g)=a(t)F(S_g(t)) + b(t) - r_gF_g(\tau,\tau_1,\tau_2)$. The above valuation is reduced to
\begin{equation}
\label{Eq:General4}
V(t)=
e^{-r_f(t-\tau)}\EE\Big[\Big(\displaystyle \widehat{F}_g(\tau,\tau_1,\tau_2)  - K\Big)^{+}\mid \mathcal{F}_t \Big],
\quad \forall t<\tau.
\end{equation}
The valuation of equation \eqref{Eq:General4} is identifiable to the valuation of a simple option where the underlying asset is a single diffusion process. 

\subsubsection{Closed formulae approximation without jumps}
The following proposition gives closed formula of the pricing depending on the parameters $a(t)$ and $b(t)$ when the gas price is modeled as a simple diffusion process.
 \begin{proposition}\label{closedForm1} 
Assume that there are no jumps on the diffusion process of the gas spot price. Then, we can identify pre-paid forward contracts on gas of the form
\begin{equation}
\widehat{F}_{t,\tau}^{P}(S_g)= \widehat{S}_ge^{-r_f(\tau-t)} \quad \mbox{and} \quad 
\widehat{F}_{t,\tau}^{P}(K)= Ke^{-r_f(\tau-t)},
\label{ChangeVariable1}
\end{equation}
where $\widehat{S}_g=\displaystyle  a(t)F(S_g(t)) + b(t) - \frac{r_g}{\tau_2 - \tau_1}\sum_{t=\tau_1}^{\tau_2}e^{-r_f}S_g(t)$. The Black-Scholes value of the spark spread option is
\begin{equation} 
V(t)=\widehat{F}_{t,\tau}^{P}(S_g) \Phi(d_1) - \widehat{F}_{t,\tau}^{P}(K) \Phi(d_2), \quad \forall t<\tau,
\label{LinearApproximation}
\end{equation}
where $d_1 =\displaystyle \frac{\ln\Big( \widehat{F}_{t,\tau}^{P}(S_g)/\widehat{F}_{t,\tau}^{P}(K) \Big) + \frac{1}{2}\sigma^2_g (\tau - t)}{\sigma_g \sqrt{\tau - t}}$ and $d_2=d_1 - \sigma_g \sqrt{\tau - t}.$

 \end{proposition}
\begin{Proof}
The proof results from the original Black-Schools closed formulae combined with the change variables as in~\eqref{ChangeVariable1}. 
\end{Proof}
 
\subsubsection{Closed formulae approximation with Merton's jumps diffusions}
The following proposition gives closed formula of the pricing depending on the parameters $a(t)$ and $b(t)$ when the gas price is modeled as a Merton's jump diffusion process.
 \begin{proposition}\label{closedForm2}
Assume that  the spot  price of gas follows a Merton's jump diffusion process as specified in~\eqref{MertonJump}.
The closed formulae of the spark spread option is
\begin{equation}
V(t)=\displaystyle e^{\lambda_2 (\tau-t)}\sum_{n=1}^{\infty} \widetilde{V}(t) \frac{(\lambda_2 (\tau - t))^n}{n!}.
\end{equation} 
$\widetilde{V}(t)$ corresponds to the pricing in~\eqref{LinearApproximation} where   $S_g$ becomes $S_g^n$ and $\sigma_g$ becomes $\sigma_g^n$:
\begin{equation}
S_g^n=S_g(0)e^{-\lambda_2 \kappa (\tau - t) + m_g n + \frac{1}{2}\delta^2}\quad \mbox{and}\quad
\sigma_g^n=\sqrt{\sigma_g^2 + \frac{n\delta^2}{\tau -t}}.
\end{equation}
 \end{proposition}
\begin{Proof} Combine the change variable in~\eqref{ChangeVariable1} and  the steps detailed in~\cite{Matsuda2004}.
\end{Proof}

\section{Conclusion}
\label{sec:Concl}
This paper outlines the challenges related to valuing gas fired power plant through spark spread options, when using different types of diffusions for the underlying processes. We have shown that  price approximations can be obtained 
using existing results. If we assume linear dependance between electricity and gas spot prices, we can further simplify the approximation by providing closed formulae easier to implement numerically. These results could be investigated through numerical simualtions in the Alberta Power Pool market, where sufficient historical data are available.

\bibliographystyle{gENO}

\bibliography{Seck_Anas}

\end{document}